\documentstyle[prd,preprint,aps,floats,eqsecnum,%
tighten,%
amssymb,amsfonts,newlfont,graphicx]{revtex}
\begin{document}
\draft
\preprint{
\begin{tabular}{r}
   DFPD 00/EP/02
\\ DFTT 03/00
\\ arXiv:hep-ex/0002020
\end{tabular}
}
\title{The Physical Significance of Confidence Intervals}
\author{Carlo Giunti}
\address{INFN, Sez. di Torino, and Dip. di Fisica Teorica,
Univ. di Torino, I--10125 Torino, Italy}
\author{Marco Laveder}
\address{Dip. di Fisica ``G. Galilei'', Univ. di Padova,
and INFN, Sez. di Padova, I--35131 Padova, Italy}
\date{20 February 2001}
\maketitle
\begin{abstract}
We define some appropriate statistical quantities
that indicate the physical significance
(reliability)
of confidence intervals
in the framework of both Frequentist and Bayesian
statistical theories.
We consider the expectation value
of the upper limit
in the absence of a signal
(that we propose to call ``exclusion potential'',
instead of ``sensitivity'' as done by Feldman and Cousins)
and
its standard deviation,
we define the ``Pull'' of a null result,
expressing the reliability of an experimental upper limit,
and
the ``upper and lower detection functions'',
that give information on the possible outcome of an experiment
if there is a signal.
We also give a new appropriate definition of ``sensitivity'',
that quantifies the capability of
an experiment to reveal the signal that is searched for
at the given confidence level.
\end{abstract}

\section{Introduction}
\label{Introduction}

The possibility to apply successfully Frequentist statistics
to problematic cases in frontier research has received
a fundamental contribution with the proposal of the Unified Approach
by Feldman and Cousins
\cite{Feldman-Cousins-98}.
The Unified Approach uses Neyman's method
\cite{Neyman-37}
to construct a confidence belt that guarantees
the derivation of confidence intervals
with a correct Frequentist coverage
(see \cite{Eadie-71,Kendall-2A,Cousins-95}).
Furthermore,
the Unified Approach allows to obtain an automatic
transition from two-sided confidence intervals to upper limits
in the case of negative results,
preserving the property of a correct Frequentist coverage.
Several works
\cite{Giunti-bo-99,Ciampolillo-98,Roe-Woodroofe-99,%
Mandelkern-Schultz-99,Punzi-99}
have followed the Unified Approach paper
discussing alternative methods
to construct the confidence belt.
Hence,
at present several Frequentist methods
with interesting properties are
available and under discussion.
Their performances are also often compared
with those of the Bayesian Theory
(see Ref.~\cite{CLW-2000}).

In this paper we define some appropriate statistical quantities
that could help to decide
which is the most appropriate method
to use in order to obtain confidence intervals
with the desired level of physical significance
(see also Ref.~\cite{Giunti-Laveder-power-00}).
We also define the sensitivity of an experiment to a signal.

The physical significance of a confidence interval
is its degree of reliability,
which is very important,
because scientists use confidence intervals
provided by experiments
or specialized compilations
(for example,
the Review of Particle Physics \cite{PDG-98})
as inputs for their calculations.
Unreliable confidence intervals
may lead to wrong or hazardous conclusions.

In Section~\ref{Exclusion}
we consider
the expectation value
of the upper limit in absence of a signal
(that we propose to call \emph{``exclusion potential''})
and its standard deviation.
In Sections~\ref{Detection} and \ref{Sensitivity}
we discuss the possibility to calculate
the expectation value
(\emph{``upper and lower detection functions''})
and the standard deviation
of the upper and lower limits
of the confidence interval produced by an experiment
in the presence of a signal
and we propose
an appropriate definition of
the \emph{``sensitivity''} of an experiment to the signal searched for.

In this article
we consider explicitly
the case of a Poisson process with known background
in the framework of Frequentist and Bayesian
statistical theories,
but similar considerations apply also to the case
of a Gaussian distribution with boundary.

\section{Exclusion potential}
\label{Exclusion}

An important quantity introduced by Feldman and Cousins
\cite{Feldman-Cousins-98}
is the average upper limit for the signal $\mu$ that would be obtained
by an ensemble of experiments if $\mu=0$
(in the case of a Poisson process with known background,
``the average upper limit that would be obtained
by an ensemble of experiments with the expected background
and no true signal'').
They called this quantity \emph{``sensitivity''},
but we think that this name is quite misleading,
because it gives the impression that this quantity
represents the expected capability of the experiment to
reveal a true signal
\footnote{
For example,
one can check on the on-line Webster Dictionary at
http://www.m-w.com/dictionary
that ``sensitivity'' is
``the quality or state of being sensitive''
and the adjective ``sensitive'' means
``capable of being stimulated or excited by external agents
(as light, gravity, or contact)''.
In our case,
since the background is known, it can be considered an
``internal agent'',
whereas the true signal is the ``external agent''
under investigation.
From the physical point of view the sensitivity
is related in many fields of application to the concept of
minimum detectable signal, therefore it is a quantity
defined for $\mu>0$, not when $\mu=0$.
}.
Instead,
the true signal is assumed to be absent.
Hence,
it is clear that the quantity under consideration
does not represent the sensitivity
of the experiment to the signal that is searched for,
but
it represents the expected upper limit for $\mu$
that will be obtained if there is no signal.
Therefore,
we propose to call this quantity
\emph{``exclusion potential''}\footnote{
The on-line Webster Dictionary at
http://www.m-w.com/dictionary
says that ``potential'' is
``something that can develop or become actual''.
},
a name that we will use in the following.
As a further justification of this name,
we note that in the case of neutrino oscillation experiments
the exclusion potential
is associated with the so-called ``exclusion curves''
in the space of the neutrino mixing parameters.

In the case of a Poisson process with known background
the exclusion potential is given by
\begin{equation}
\mu_{\mathrm{ep}}(b,\alpha)
=
\sum_{n=0}^{\infty}
\mu_{\mathrm{up}}(n,b,\alpha)
\,
P(n|\mu=0,b)
\,,
\label{ep}
\end{equation}
where
$n$ is the number of counts,
$b$ is the expected mean background,
$\mu$ is the mean true signal,
$P(n|\mu,b)$ is the Poisson p.d.f. for the process
and
$\mu_{\mathrm{up}}(n,b,\alpha)$
is the upper limit
of the $100(1-\alpha)\%$ confidence interval
for $\mu$
corresponding to $n$ counts.
The exclusion potential $\mu_{\mathrm{ep}}$
depends on the values of the upper limits $\mu_{\mathrm{up}}(n,b,\alpha)$,
which are different in the different methods
for calculating the confidence belt.

It is interesting to note that the above definition of exclusion potential
can be extended to the results obtained with the Bayesian Theory
(see, for example, \cite{DAgostini-YR3-99})
if $\mu_{\mathrm{up}}$
is interpreted as the upper limit of the Bayesian credibility interval
for $\mu$.
In the following we will consider
the Bayesian Theory
assuming a flat prior
and shortest credibility intervals for $\mu$.
In this case the posterior p.d.f. for $\mu$
is
\begin{equation}
P(\mu|n,b)
=
( b + \mu )^n \, e^{-\mu}
\left( \displaystyle n! \, \sum_{k=0}^{n} \frac{b^k}{k!} \right)^{-1}
\,,
\label{posterior}
\end{equation}
and the probability
(degree of belief)
that the true value of $\mu$ lies in the range
$[\mu_1,\mu_2]$
is given by
\begin{equation}
P(\mu\in[\mu_1,\mu_2]|n,b)
=
\left(
e^{-\mu_1} \sum_{k=0}^{n} \frac{(b+\mu_1)^k}{k!}
-
e^{-\mu_2} \sum_{k=0}^{n} \frac{(b+\mu_2)^k}{k!}
\right)
\left( \displaystyle \sum_{k=0}^{n} \frac{b^k}{k!} \right)^{-1}
\,.
\label{integral-probability}
\end{equation}
The shortest $100(1-\alpha)\%$ credibility intervals
$[\mu_{\mathrm{low}},\mu_{\mathrm{up}}]$
are obtained by choosing
$\mu_{\mathrm{low}}$
and
$\mu_{\mathrm{up}}$
such that
$P(\mu\in[\mu_{\mathrm{low}},\mu_{\mathrm{up}}]|n,b)=1-\alpha$
and
$P(\mu_{\mathrm{low}}|n,b) = P(\mu_{\mathrm{up}}|n,b)$
if possible
(with $\mu_{\mathrm{low}} \geq 0$),
or
$\mu_{\mathrm{low}} = 0$.

\begin{figure}
\begin{center}
\mbox{\includegraphics[bb=56 105 525 800,height=0.8\textheight]{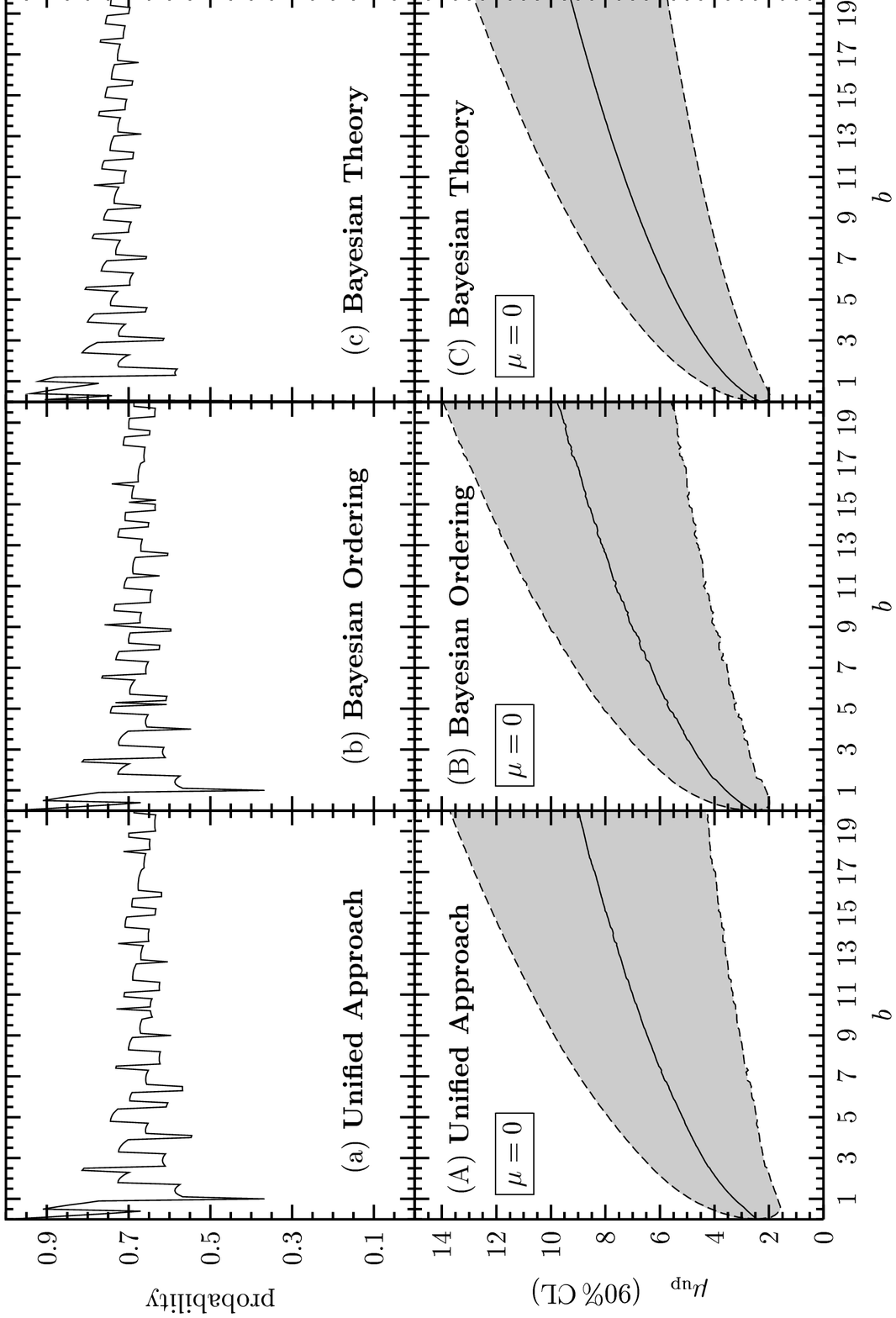}}
\end{center}
\caption{ \label{exposigma}
90\% CL exclusion potential $\mu_{\mathrm{ep}}$ (solid line)
and
$\mu_{\mathrm{ep}} \pm \sigma_{\mathrm{ep}}$
(dashed lines)
as functions of the known background $b$
obtained with
the Unified Approach (A) \protect\cite{Feldman-Cousins-98},
with
the Bayesian Ordering method (B) \protect\cite{Giunti-bo-99}
and with
the Bayesian Theory (C) with a flat prior
and shortest credibility intervals.
The probability to obtain a 90\% CL upper bound
in the shadowed region if $\mu=0$ is shown
in the three upper figures
a, b and c.
}
\end{figure}

The solid lines in Figs.~\ref{exposigma} and \ref{exporange}
represent the exclusion potential as a function of the background
in the interval
$0 \leq b \leq 20 $
for a confidence level $90\%$ ($\alpha=0.10$)
obtained with
the Unified Approach \cite{Feldman-Cousins-98}
(Figs.~\ref{exposigma}A and \ref{exporange}A),
with the Bayesian Ordering method\footnote{
The Bayesian Ordering method,
as the Unified Approach,
is a Frequentist method with correct coverage
and
automatic
transition from two-sided confidence intervals to upper limits
in the case of negative results.
} \cite{Giunti-bo-99}
(Figs.~\ref{exposigma}B and \ref{exporange}B) and
with the Bayesian Theory
assuming a flat prior
and shortest credibility intervals
(Figs.~\ref{exposigma}C and \ref{exporange}C).

Feldman and Cousins
\cite{Feldman-Cousins-98}
suggested that
in the cases in which the measurement is less than
the estimated mean background and a stringent upper bound on $\mu$
is inferred,
the experimental collaboration should report
also the exclusion potential
(that they call ``sensitivity'') of the experiment.
This is also recommended by the Particle Data Group
\cite{PDG-98}.

In practice,
the comparison of the upper bound and the exclusion potential
is used as an assessment of the reliability
of the upper bound.
However, one can notice that
a simple comparison of the
upper bound with the exclusion potential
does not give any information unless
a meaningful scale of comparison is given.
We think that a meaningful scale is the
possible fluctuation of the
upper bound $\mu_{\mathrm{up}}(n,b,\alpha)$
in an ensemble of experiments with the expected background
and no true signal.
A quantification of this fluctuation is provided by the
standard deviation $\sigma_{\mathrm{ep}}(b,\alpha)$
of the upper limit $\mu_{\mathrm{up}}$
calculated assuming $\mu=0$,
\begin{equation}
\sigma_{\mathrm{ep}}^2(b,\alpha)
=
\sum_{n=0}^{\infty}
\left[ \mu_{\mathrm{up}}(n,b,\alpha) - \mu_{\mathrm{ep}}(b,\alpha) \right]^2
P(n|\mu=0,b)
\,.
\label{sigmas}
\end{equation}
The shadowed regions delimited by the dashed lines
in Fig.~\ref{exposigma}
represent the range
$[
\mu_{\mathrm{ep}}-\sigma_{\mathrm{ep}}
,
\mu_{\mathrm{ep}}+\sigma_{\mathrm{ep}}
]$
for $\alpha=0.10$
obtained with
the Unified Approach (A),
with the Bayesian Ordering method (B)
and
with the Bayesian Theory
assuming a flat prior
and shortest credibility intervals.
The corresponding probability
to obtain an upper bound
in the interval
$[
\mu_{\mathrm{ep}}-\sigma_{\mathrm{ep}}
,
\mu_{\mathrm{ep}}+\sigma_{\mathrm{ep}}
]$
if $\mu=0$ is shown
in the three upper figures
\ref{exposigma}a, \ref{exposigma}b and \ref{exposigma}c.
One can see that,
except for small values of the background $b$,
the probability
to obtain an upper bound
in the interval
$[
\mu_{\mathrm{ep}}-\sigma_{\mathrm{ep}}
,
\mu_{\mathrm{ep}}+\sigma_{\mathrm{ep}}
]$
if $\mu=0$
is not far from 68\%
in the three considered methods.
The probability curves in Fig.~\ref{exposigma} have wild jumps
because $n$ is an integer and $\mu_{\mathrm{up}}$
has discrete jumps as $n$ is varied.

\begin{figure}
\begin{center}
\mbox{\includegraphics[bb=264 108 525 800,height=0.8\textheight]{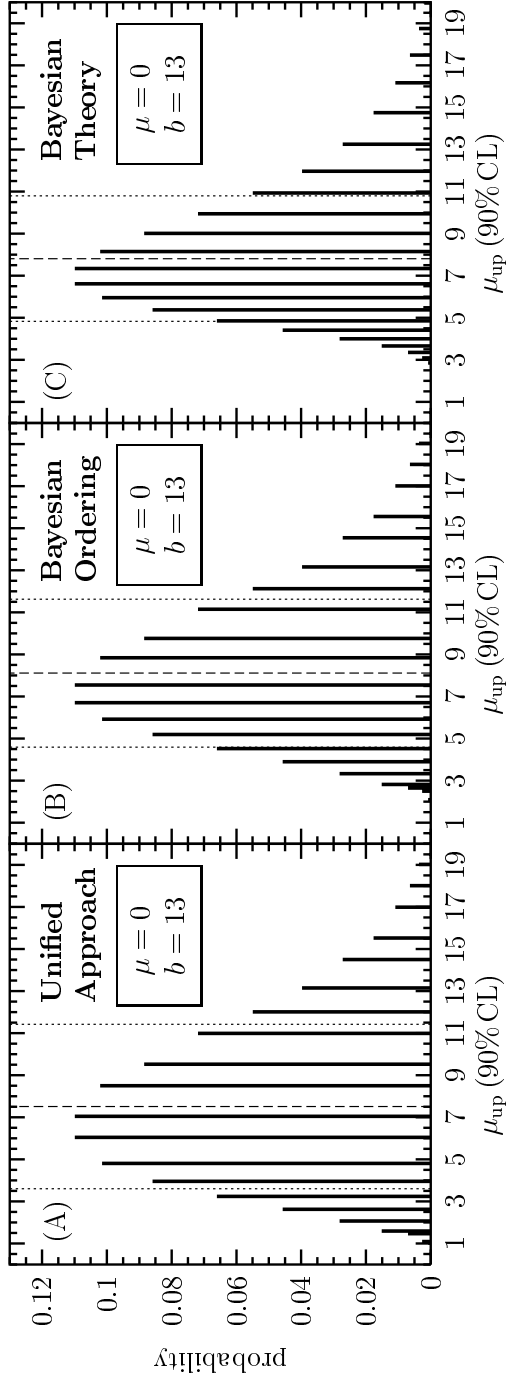}}
\end{center}
\caption{ \label{muupprob}
Probability of the possible values of the 90\% CL upper bound
$\mu_{\mathrm{up}}$
in the absence of a signal ($\mu=0$) and
a known background $b=13$
in the Unified Approach (A) \protect\cite{Feldman-Cousins-98},
in the Bayesian Ordering method (B) \protect\cite{Giunti-bo-99}
and
in the Bayesian Theory (C) with a flat prior
and shortest credibility intervals.
The dashed vertical lines show the value of
the exclusion potential
$\mu_{\mathrm{ep}}$
and the dotted vertical lines represent
$\mu_{\mathrm{ep}} \pm \sigma_{\mathrm{ep}}$.
}
\end{figure}

\begin{figure}
\begin{center}
\mbox{\includegraphics[bb=264 108 525 800,height=0.8\textheight]{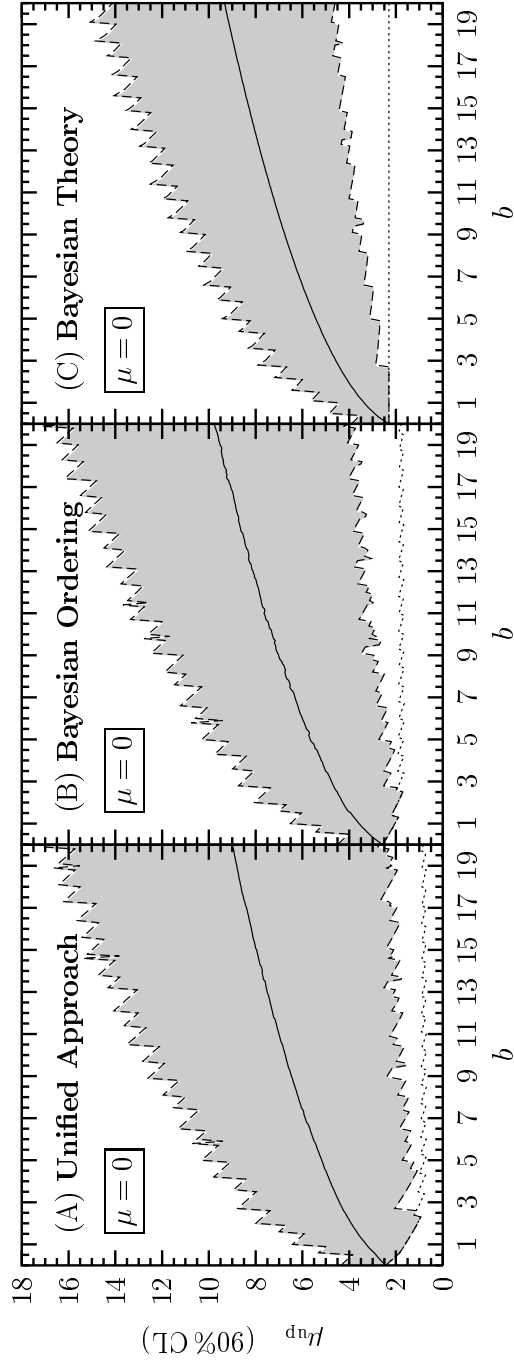}}
\end{center}
\caption{ \label{exporange}
90\% CL exclusion potential $\mu_{\mathrm{ep}}$ (solid line)
and
90\% width of the 90\% CL upper limit
(shadowed region delimited by the dashed lines)
obtained with
the Unified Approach (A) \protect\cite{Feldman-Cousins-98},
with
the Bayesian Ordering method (B) \protect\cite{Giunti-bo-99}
and with
the Bayesian Theory (C) with a flat prior
and shortest credibility intervals.
The dotted lines represent the lower limit for $\mu_{\mathrm{up}}$(90\% CL),
obtained for $n=0$.
}
\end{figure}

As an illustration,
in Fig.~\ref{muupprob}
we have plotted the 
probability of the possible values of the 90\% CL upper bound
$\mu_{\mathrm{up}}$
if $\mu=0$ and $b=13$.
One can see that the upper bound
$\mu_{\mathrm{up}}$
can assume only discrete values.
The probability
to obtain an upper bound
in the interval
$[
\mu_{\mathrm{ep}}-\sigma_{\mathrm{ep}}
,
\mu_{\mathrm{ep}}+\sigma_{\mathrm{ep}}
]$,
delimited by the dotted vertical lines
in Fig.~\ref{muupprob},
has discrete jumps as $b$ is changed,
depending on which possible values of the upper bound
are included in the interval.

It is also possible to calculate
the possible range of fluctuation of the upper bound
$\mu_{\mathrm{up}}$
with a desired probability $\gamma$.
The shadowed regions delimited by the dashed lines
in Fig.~\ref{exporange}
show the 90\% width of the 90\% CL upper limit,
\textit{i.e.}
the possible range of fluctuation of the
$\mu_{\mathrm{up}}(n,b,\alpha=0.90)$
with probability $\gamma=0.90$
as a function of $b$ if $\mu=0$.
This range of fluctuation has been calculated
in order to obtain a band as symmetric as possible
around $\mu_{\mathrm{ep}}$.
In practice this is done by a computer program that simultaneously
decreases the lower limit $\mu_{\mathrm{up}}^{(-)}(b,\alpha,\gamma)$
of the band
(lower dashed lines in Fig.~\ref{exporange})
and
increases the upper limit $\mu_{\mathrm{up}}^{(+)}(b,\alpha,\gamma)$
of the band
(upper dashed lines in Fig.~\ref{exporange})
until the condition
\begin{equation}
\sum_{n=n^{(-)}(b,\alpha,\gamma)}^{n^{(+)}(b,\alpha,\gamma)}
P(n|\mu=0,b) \geq \gamma
\label{band}
\end{equation}
is reached.
Here
$n^{(-)}(b,\alpha,\gamma)$
and
$n^{(+)}(b,\alpha,\gamma)$
are the values of $n$ such that
\begin{equation}
\mu_{\mathrm{up}}(n^{(-)}(b,\alpha,\gamma),b,\alpha)
=
\mu_{\mathrm{up}}^{(-)}(b,\alpha,\gamma)
\,,
\quad
\mu_{\mathrm{up}}(n^{(+)}(b,\alpha,\gamma),b,\alpha)
=
\mu_{\mathrm{up}}^{(+)}(b,\alpha,\gamma)
\,.
\label{n_pm}
\end{equation}
The inequality sign in Eq.~(\ref{band}) is needed
because $n$ is an integer and in general it is not possible to obtain
exactly the desired probability $\gamma$.
This is also the reason for the fact that
the dashed lines in Fig.~\ref{exporange} are not smooth.
The dotted lines in Fig.~\ref{exporange}
represent the lower limit for
$\mu_{\mathrm{up}}(n,b,\alpha=0.90)$
as a function of $b$,
that is obtained for $n=0$.

It is clear that
having analyzed the data using,
for example, the Unified Approach and having obtained an upper bound
$\mu_{\mathrm{up}}$
with an expected background $b$,
looking at Fig.~\ref{exporange}A
one can judge if the upper bound
$\mu_{\mathrm{up}}$
is reasonable or the result of an unlikely statistical fluctuation.
For example,
the Heidelberg-Moscow
double-beta decay experiment \cite{Baudis-99} measured recently
$n=7$ events with an expected background $b=13$.
Using the Unified Approach they obtained
$\mu_{\mathrm{up}} = 2.07$ at 90\% CL,
with exclusion potential $\mu_{\mathrm{ep}} = 7.51$.
Looking at Fig.~\ref{exporange}A one can see that
the 90\% lower limit for $\mu_{\mathrm{up}}$ assuming
$\mu=0$
is $\mu_{\mathrm{up}}^{(-)} = 2.07$,
so the discrepancy between $\mu_{\mathrm{up}}$ and $\mu_{\mathrm{ep}}$
is just acceptable at the border of 10\% probability.
Using the Bayesian Ordering method
they would have obtained
$\mu_{\mathrm{up}}(n=7,b=13,\alpha=0.10) = 3.33$,
with exclusion potential
$\mu_{\mathrm{ep}}(b=13,\alpha=0.10) = 8.11$,
and with the Bayesian Theory
(with a flat prior and shortest credibility intervals)
$\mu_{\mathrm{up}}(n=7,b=13,\alpha=0.10) = 4.01$
and
$\mu_{\mathrm{ep}}(b=13,\alpha=0.10) = 7.81$.

Of course,
one can obtain the same results using the measured value of $n$
and the Poisson p.d.f. $P(n|\mu=0,b)$ with the expected background $b$.
The advantage of Figs.~\ref{exposigma} and \ref{exporange}
is that one can easily transform $\mu_{\mathrm{up}}$ and $\mu_{\mathrm{ep}}$
in the physical quantity of interest,
presenting the result only in terms of physical quantities.
For example,
in double-beta decay experiments
$\mu$ is connected to the effective Majorana mass
$|\langle{m}\rangle|$
of the electron neutrino by the relation
\begin{equation}
|\langle{m}\rangle|
=
\xi
\,
\frac{\sqrt{\mu}}{|\mathcal{M}|}
\,,
\label{bb}
\end{equation}
where
$\xi$
is a constant with dimension of mass
that depends on the decaying nucleus
and
$\mathcal{M}$
is the nuclear matrix element.
In the case of the Heidelberg-Moscow
double-beta decay experiment \cite{Baudis-99},
the decaying nucleus is $^{76}$Ge, with
$\xi = 0.57 \, \mathrm{eV}$.
Using the nuclear matrix element
$|\mathcal{M}| = 2.80$
calculated in \cite{Simkovic-99},
the 90\% CL upper bound for the effective Majorana mass
obtained with the Unified Approach is
$ |\langle{m}\rangle|_{\mathrm{up}} = 0.29 \, \mathrm{eV} $,
the corresponding exclusion potential is
$ |\langle{m}\rangle|_{\mathrm{ep}} = 0.56 \, \mathrm{eV} $
and the
90\% lower limit for
the fluctuations of $|\langle{m}\rangle|_{\mathrm{up}}$
(if
$|\langle{m}\rangle|=0$)
is
$ |\langle{m}\rangle|_{\mathrm{up}}^{(-)} = 0.29 \, \mathrm{eV} $.
Using instead the Bayesian Ordering method,
we obtain
$ |\langle{m}\rangle|_{\mathrm{up}} = 0.37 \, \mathrm{eV} $,
$ |\langle{m}\rangle|_{\mathrm{ep}} = 0.58 \, \mathrm{eV} $
and
$ |\langle{m}\rangle|_{\mathrm{up}}^{(-)} = 0.37 \, \mathrm{eV} $,
and with the Bayesian Theory
(with a flat prior and shortest credibility intervals)
we have
$ |\langle{m}\rangle|_{\mathrm{up}} = 0.41 \, \mathrm{eV} $,
$ |\langle{m}\rangle|_{\mathrm{ep}} = 0.57 \, \mathrm{eV} $
and
$ |\langle{m}\rangle|_{\mathrm{up}}^{(-)} = 0.41 \, \mathrm{eV} $.

Let us define the Pull of a null result as
\begin{equation}
\mathrm{Pull}(n,b,\alpha)
=
\frac{
\mu_{\mathrm{up}}(n,b,\alpha)
-
\mu_{\mathrm{ep}}(b,\alpha)
}
{ \sigma_{\mathrm{ep}}(b,\alpha) }
\,.
\label{pull}
\end{equation}
If $\mathrm{Pull}(n,b,\alpha) \gtrsim 1$,
the experimental upper limit
$\mu_{\mathrm{up}}$
is significantly weaker than the exclusion potential
$\mu_{\mathrm{ep}}$
and may be considered as a weak indication that a signal may be present
($\mu>0$).
On the other hand,
if $\mathrm{Pull}(n,b,\alpha) \lesssim -1$
and there is no doubt on the value of the mean background $b$,
it means that
the experiment has experienced an unlikely low fluctuation of the
background and the resulting upper bound,
that is significantly more stringent than the exclusion potential,
is not reliable
from a physical point of view
and it is likely to increase if the experiment is continued
(a quite undesirable behavior),
as shown by the example of the KARMEN experiment
(see Fig.~4 of Ref.~\cite{KARMEN-Moriond99}).
Moreover,
a method that gives values of the Pull closer to zero
produces upper bounds that are more reliable from a physical point of view.

For example,
in the case of the Heidelberg-Moscow
double-beta decay experiment \cite{Baudis-99},
we have
$n=7$,
$b=13$
and
$\alpha=0.10$,
giving
$\sigma_{\mathrm{ep}} = 3.91$
and
$\mathrm{Pull} = -1.39$
in the Unified Approach,
$\sigma_{\mathrm{ep}} = 3.52$
and
$\mathrm{Pull} = -1.36$
with the Bayesian Ordering method,
and
$\sigma_{\mathrm{ep}} = 2.98$
and
$\mathrm{Pull} = -1.27$
in the Bayesian Theory
with a flat prior and shortest credibility intervals.
Hence,
among the two Frequentist methods that we have considered,
the upper bound obtained with Bayesian Ordering
is slightly more reliable,
from a physical point of view,
than the one obtained with the Unified Approach.
If one is willing to accept the Bayesian Theory,
the corresponding upper bound is clearly the
most reliable one from a physical point of view.

Let us point out that
the knowledge of the possible fluctuation of
the upper bound
$\mu_{\mathrm{up}}$
with respect to the exclusion potential $\mu_{\mathrm{ep}}$
can also help to
\emph{decide,
before looking at the data,
which is the most appropriate Frequentist method
for the statistical analysis}.

This can be understood
comparing, for example, Figs.~\ref{exporange}A and \ref{exporange}B,
obtained with two Frequentist methods with correct coverage.
One can see that the upper dashed
lines in the two figures almost coincide
and
the exclusion potential is slightly lower in the Unified Approach,
with a difference 
going from 0.30 for $b=1$
to 0.73 for $b=19$,
with a relative difference of 8--9\%.
On the other hand,
the lower dashed lines and the dotted lines
obtained with the Bayesian Ordering are
significantly higher than those obtained with the Unified Approach.
The difference of the lower dashed lines,
goes from 0.43 for $b=1$
to 1.40 for $b=19$,
with a relative difference going from 27\% to 64\%.
The difference between the lowest possible values
of $\mu_{\mathrm{up}}$ (dotted lines)
is quite large:
for $b \gtrsim 4$
the lowest possible values
of $\mu_{\mathrm{up}}$
is about 0.8
in the Unified Approach
and about 1.8
in the Bayesian Ordering method,
more than twice!

Hence,
if one does not want to risk
to have to present a very stringent limit,
which would be statistically correct but physically misleading,
in the case of observation of less events than the expected background,
one can look at figures like
Figs.~\ref{exporange}A and \ref{exporange}B
and the corresponding ones for other Frequentist methods
and decide which is the method more suitable for his tastes.
Let us emphasize that this choice
\emph{must be done before looking at the data}.
If one chooses the statistical method on the basis of the data,
the property of coverage is lost.

Furthermore,
the exclusion potential of an experiment can be calculated
and published
before starting the experiment or before the data are known,
in order to have an indication of the excluded region that will be obtained
in the case of absence of a signal
(or a signal much smaller than the expected background).
We think that
it would be useful to publish
together with the exclusion potential
also the standard deviation of the upper limit in the absence of a signal,
in order to illustrate the possible fluctuations of the excluded region
and to give,
at the same time, a quantitative statement on the precision
of the experiment.

\section{Detection functions}
\label{Detection}

\begin{figure}
\begin{center}
\mbox{\includegraphics[bb=56 105 525 800,height=0.78\textheight]{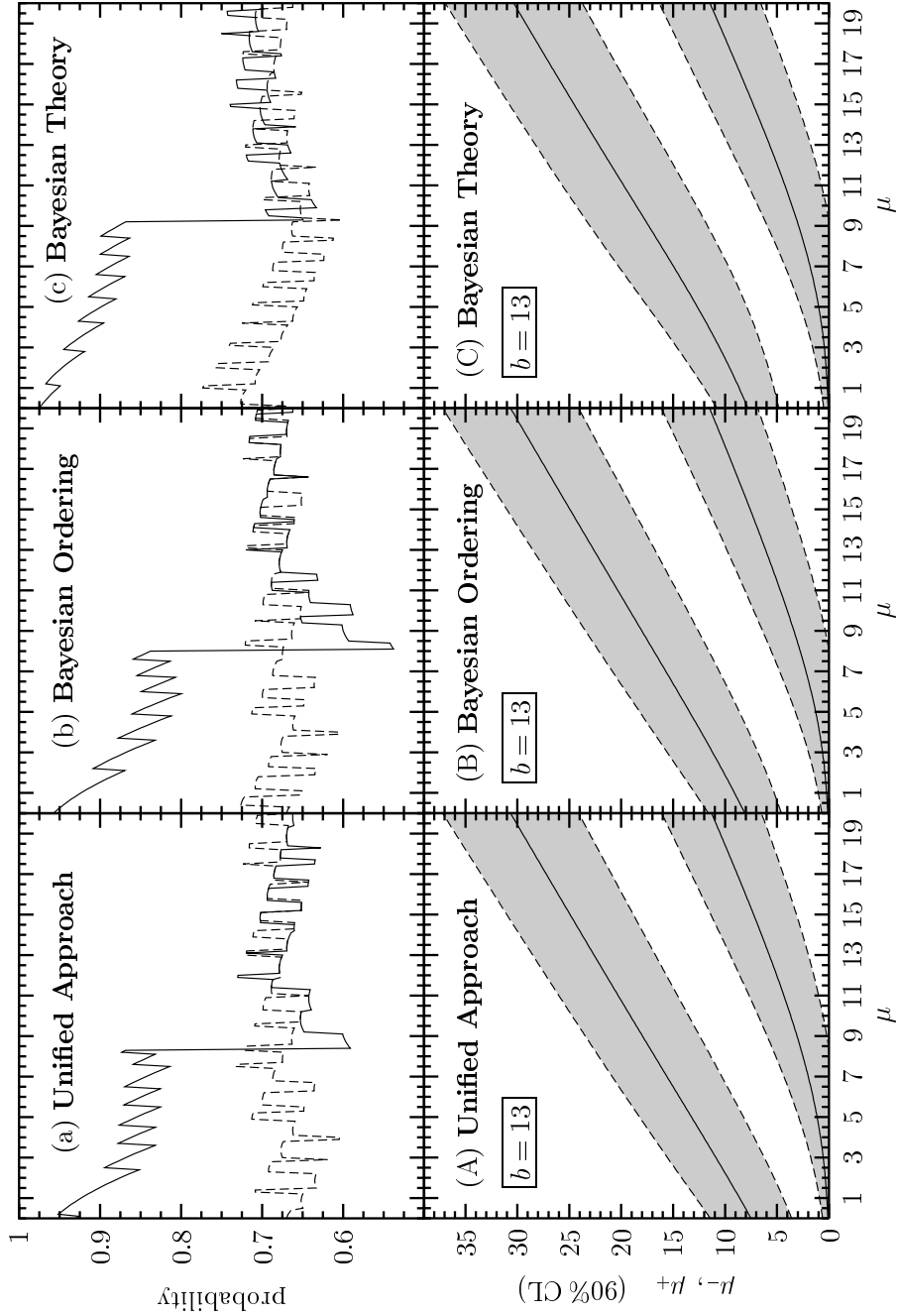}}
\end{center}
\caption{ \label{mu}
Upper and lower 90\% CL detection functions ($\alpha=0.10$)
(Eq.~(\ref{detection}))
$\mu_{+}$ (upper solid lines) and $\mu_{-}$ (lower solid lines)
as functions of $\mu$ for $b=13$
obtained with
the Unified Approach \protect\cite{Feldman-Cousins-98}
(A),
with the Bayesian Ordering method \protect\cite{Giunti-bo-99}
(B) and
with the Bayesian Theory
assuming a flat prior
and shortest credibility intervals
(C).
The shadowed regions delimited by the dashed lines
represent the bands
$\mu_{+} \pm \sigma_{+}$
(upper band)
and
$\mu_{-} \pm \sigma_{-}$
(lower band),
with $\sigma_{\pm}$ defined in Eq.~(\ref{sigma-detection}).
The three upper figures a, b and c
show the probability to find an upper limit
$\mu_{\mathrm{up}}$
in the interval
$\mu_{+} \pm \sigma_{+}$
(dashed line)
and
the probability to find an lower limit
$\mu_{\mathrm{low}}$
in the interval
$\mu_{-} \pm \sigma_{-}$
(solid line).
}
\end{figure}

The exclusion potential and
the standard deviation of the upper limit in the absence of a signal
are interesting quantities,
but they give only information on the possible experimental result
in the worst-case scenario,
that in which the signal is absent or so small to be undetectable.
Usually
researchers are more interested in finding positive signals.
For example,
they would like to know in advance which would be
the most likely outcome of the experiment
if there is a true signal.
In this case,
we propose to calculate the
\emph{upper and lower detection functions},
$\mu_{+}(\mu,b,\alpha)$
and
$\mu_{-}(\mu,b,\alpha)$,
obtained averaging the upper and lower limits
$\mu_{\mathrm{up}}(n,b,\alpha)$
and
$\mu_{\mathrm{low}}(n,b,\alpha)$
over $n$ with the Poisson p.d.f. $P(n|\mu,b)$:
\begin{equation}
\mu_{\pm}(\mu,b,\alpha)
=
\sum_{n=0}^{\infty}
\mu_{\mathrm{up}\atop\mathrm{low}}(n,b,\alpha)
\,
P(n|\mu,b)
\,.
\label{detection}
\end{equation}
The standard deviation of $\mu_{\mathrm{up}\atop\mathrm{low}}(n,b,\alpha)$
is given by
\begin{equation}
\sigma_{\pm}^2(\mu,b,\alpha)
=
\sum_{n=0}^{\infty}
\left[
\mu_{\mathrm{up}\atop\mathrm{low}}(n,b,\alpha)
-
\mu_{\pm}(\mu,b,\alpha)
\right]^2
\,
P(n|\mu,b)
\,.
\label{sigma-detection}
\end{equation}
In Fig.~\ref{mu}
we have plotted $\mu_{+}$, $\mu_{-}$
(upper and lower solid lines, respectively)
as functions of $\mu$ for $\alpha=0.10$ and $b=13$
in the Unified Approach \cite{Feldman-Cousins-98}
(Fig.~\ref{mu}A),
in the Bayesian Ordering method \cite{Giunti-bo-99}
(Fig.~\ref{mu}B) and
in the Bayesian Theory
assuming a flat prior
and shortest credibility intervals
(Fig.~\ref{mu}C).
The shadowed regions delimited by the dashed lines
in Fig.~\ref{mu}
represent the bands
$\mu_{+} \pm \sigma_{+}$
(upper band)
and
$\mu_{-} \pm \sigma_{-}$
(lower band).
From Fig.~\ref{mu}
one can see that the average upper bound $\mu_{+}$
is almost identical in the three methods,
but the range $\mu_{+} \pm \sigma_{+}$
for small values of $\mu$
is shortest in the Bayesian Theory
and largest in the Unified Approach.
The average lower bound $\mu_{-}$
and the range $\mu_{-} \pm \sigma_{-}$
are similar in the three methods,
with the small difference that
$\mu_{-} - \sigma_{-} > 0$
for $\mu \simeq 9$ in the two Frequentist methods
(Unified Approach and Bayesian Ordering)
and
$\mu \simeq 10$ in the Bayesian Theory.
The three upper plots in Fig.~\ref{mu}
show the probability to find
$\mu_{\mathrm{low}}$
in the interval
$\mu_{-} \pm \sigma_{-}$
(solid lines)
and
the probability to find
$\mu_{\mathrm{up}}$
in the interval
$\mu_{+} \pm \sigma_{+}$
(dashed lines).
The latter probability is high (larger than 80\%)
for small values of $\mu$,
where the interval $\mu_{-} \pm \sigma_{-}$
includes zero,
and stabilizes around 68\%
for higher values of $\mu$
(the fluctuations and discontinuities of the probability
as a function of $\mu$
are due to the discreteness of $n$).

The detection functions and the
standard deviations of the lower and upper bounds
show the
expected result and its possible fluctuations
if the signal under measurement is not negligibly small.
In the next section we present a definition of
sensitivity of an experiment to a signal.

\section{Sensitivity to a signal}
\label{Sensitivity}

Often researchers would like to plan an experiment
capable of revealing a signal whose value
is indicated by previous measurements
or predicted by theory.
Hence,
it is useful to define the
\emph{sensitivity of an experiment to a signal}.

Two probabilities must be involved
in the definition of the sensitivity to a signal:
the confidence level $1-\alpha$
of the confidence interval that represents the result of the experiment
and
the probability $\lambda$
to find a confidence interval with a positive lower bound.

We think that an appropriate definition of the
$100\lambda\%$
sensitivity corresponding to a $100(1-\alpha)\%$ confidence level,
$\mu_{\mathrm{s}}(b,\alpha,\lambda)$,
of an experiment measuring a Poisson process with known background $b$
is
\emph{the value of $\mu$ for which there is a probability
$\lambda$ to find a positive lower limit for $\mu$
with confidence level $100(1-\alpha)\%$}.
Hence,
we define $\mu_{\mathrm{s}}(b,\alpha,\lambda)$
through the equation\footnote{
After the completion of this work,
we have been informed that
Hernandez, Nava and Rebecchi \cite{Hernandez-Nava-Rebecchi-discovery-96}
defined similar criteria in order to calculate
``discovery limits'' in prospective studies.
Our Eq.~(\ref{mu_s})
coincides with their Eq.~(6) with $\delta = 1 - \lambda$,
and our Eq.~(\ref{mu_low})
corresponds to their Eq.~(5) with $\epsilon = \alpha$
in the case of
Frequentist methods with correct coverage
and automatic transition from two-sided confidence intervals to upper limits
for a small number of counts.
}
\begin{equation}
\sum_{n \geq n_{\mathrm{s}}(b,\alpha)} P(n|\mu_s(b,\alpha,\lambda),b) = \lambda
\,,
\label{mu_s}
\end{equation}
where $n_{\mathrm{s}}(b,\alpha)$ is the smallest integer such that
\begin{equation}
\mu_{\mathrm{low}}(n_{\mathrm{s}}(b,\alpha),b,\alpha) > 0
\,.
\label{mu_low}
\end{equation}
Here
$\mu_{\mathrm{low}}(n,b,\alpha)$
is the lower limit of the $100(1-\alpha)\%$ confidence interval
(credibility interval in the Bayesian Theory)
corresponding to the observation of $n$ events.
In all Frequentist methods with correct coverage
that guarantee
an automatic transition from two-sided confidence intervals to upper limits
for $n \lesssim b$
(as the Unified Approach \cite{Feldman-Cousins-98}
and the Bayesian Ordering method \cite{Giunti-bo-99}),
the acceptance interval for $\mu=0$
starts at
$n_1(\mu=0,b,\alpha)=0$
and ends at
$n_2(\mu=0,b,\alpha)$,
where
$n_2(\mu=0,b,\alpha)$
is the smallest integer such that
\begin{equation}
\sum_{n=0}^{n_2(\mu=0,b,\alpha)} P(n|\mu=0,b) \geq 1-\alpha
\,.
\label{n_2}
\end{equation}
Then it is clear that
\begin{equation}
n_{\mathrm{s}}(b,\alpha) = n_2(\mu=0,b,\alpha) + 1
\label{n_s}
\end{equation} 
is the smallest integer that satisfies Eq.~(\ref{mu_low}).

\begin{figure}
\begin{center}
\mbox{\includegraphics[bb=12 123 282 802,height=0.8\textheight]{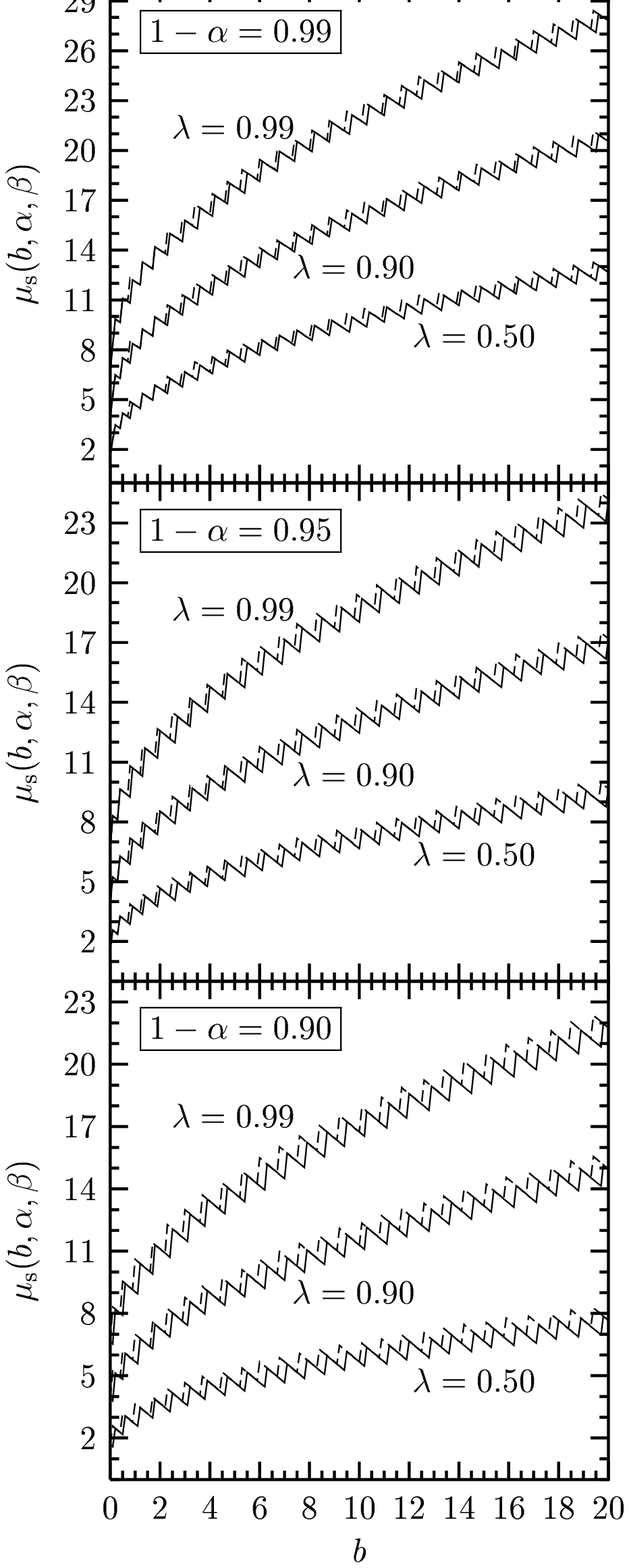}}
\end{center}
\caption{ \label{sensi}
Sensitivity
$\mu_{\mathrm{s}}(b,\alpha,\lambda)$
(see Eq.~(\ref{mu_s}))
as a function of $b$
in Frequentist methods
(solid lines)
and in the Bayesian Theory
with a flat prior
and shortest credibility interval
(dashed lines)
for $1-\alpha=0.90,\,0.95,\,0.99$ and $\lambda=0.50,\,0.90,\,0.99$.
}
\end{figure}

Figure~\ref{sensi}
shows the value of
$\mu_{\mathrm{s}}(b,\alpha,\lambda)$
as a function of $b$
in Frequentist methods
(solid lines)
and in the Bayesian Theory
with a flat prior
and shortest credibility intervals
(dashed lines)
for $1-\alpha=0.90,\,0.95,\,0.99$ and $\lambda=0.50,\,0.90,\,0.99$.
The lines are not smooth because of the discreteness of
$n_{\mathrm{s}}(b,\alpha)$
that causes jumps of the solution of Eq.~(\ref{mu_s})
as $b$ varies from one value to another with different
$n_{\mathrm{s}}(b,\alpha)$.

The sensitivity
$\mu_{\mathrm{s}}(b,\alpha,\lambda)$
provides useful information for the planning of an experiment
with the purpose of exploring a range
$[\mu_{\mathrm{min}},\mu_{\mathrm{max}}]$
of possible values of $\mu$
that could be inferred from the results of other experiments
or from theory.
In order to do this,
the background in the experiment must be small enough
that the sensitivity $\mu_{\mathrm{s}}(b,\alpha,\lambda)$
is smaller than $\mu_{\mathrm{min}}$.
In this case
the experiment will have probability $1-\alpha$
to obtain a correct result
(\textit{i.e.} a confidence interval that contains the true value of $\mu$)
and a probability bigger than $\lambda$
to obtain a positive lower limit,
\textit{i.e.} to reveal a true signal,
if $\mu>\mu_{\mathrm{s}}(b,\alpha,\lambda)$.
The probability that the experiment will reveal
a true signal
within a correct confidence interval is larger the product $(1-\alpha)\lambda$
(if $\mu>\mu_{\mathrm{s}}(b,\alpha,\lambda)$).
Therefore,
it is desirable to have both $1-\alpha$ and $\lambda$ large.
In Fig.~\ref{sensi} we have chosen $1-\alpha=0.90,\,0.95,\,0.99$,
that are commonly used values for the confidence level,
and
$\lambda=0.50,\,0.90,\,0.99$.
We think that
$1-\alpha$
should be always chosen large,
preferably $1-\alpha=0.99$,
because getting a correct result is the most important thing.
As for $\lambda$,
a large value is important in order to have good chances to reveal the signal
(if it exist!).
For example,
having $1-\alpha=0.99$ and $\lambda=0.99$
gives a probability bigger than 98\%
to find a true signal
within a correct confidence interval
(if $\mu>\mu_{\mathrm{s}}(b,\alpha,\lambda)$).
On the other hand,
for $1-\alpha=0.90$ and $\lambda=0.50$ ($\lambda=0.90$)
the probability
to find a true signal
within a correct confidence interval
can be as low as
45\% (81\%).

From Fig.~\ref{sensi}
one can see that the sensitivity increases sub-linearly
as the background increases.
Since the background increases
linearly with the time of data-taking,
the sensitivity of the experiment increases sub-linearly
as a function of data-taking time.
Let us consider an experiment searching for a signal
produced by a new process for which there is an indication
from previous experiments or from theory.
Since the signal,
as the background,
increases linearly with the time of data-taking,
there is a time such that the signal becomes larger than the sensitivity
and this time
provides an estimate
of the data-taking time
necessary to reveal the new process.

It is interesting to notice that the sensitivity
for $b=13$, $1-\alpha=0.90$
is
$\mu_{\mathrm{s}} \simeq 7$
for
$\lambda=0.50$
and
$\mu_{\mathrm{s}} \simeq 13$
for
$\lambda=0.90$,
in rough agreement with the
lower bands in Fig.~\ref{mu},
which show that there is a good chance to find a lower limit for $\mu$
bigger than zero if $\mu \gtrsim 10$.

From this example
it is clear that the sensitivity of an experiment
is different from its exclusion potential.
The proposals of new experiments on the search for a signal
should present the sensitivity as the most
interesting characteristic of the experiment.
The exclusion potential should be also presented
as an illustration of the potentiality of the experiment
in the most unfortunate case of
absence of a signal.

Let us remind the reader that the word ``probability''
has different definitions in the Frequentist and Bayesian theories.
In the Bayesian theory
``probability''
is defined as ``degree of belief'',
whereas in the Frequentist theory
it is defined as ratio of the number of positive cases and total number of trials
in a large ensemble.
The Frequentist definition avoids the need of subjective judgment,
but it is not clear what is its meaning in the planning and realization
of \emph{one} experiment (or a few experiments).
Whatever the meaning of Frequentist probability in this case,
we think that it is comforting to see from Fig.~\ref{sensi}
that the Frequentist and Bayesian values for
$\mu_{\mathrm{s}}(b,\alpha,\lambda)$
are quite close.

\section{Conclusions}
\label{Conclusions}

In conclusion,
we have defined some quantities that may help
to asses the reliability (physical significance)
of the confidence intervals obtained with different methods.
We have also defined appropriately
the sensitivity of an experiment to a signal.

In Section~\ref{Exclusion}
we have considered the quantity called
``sensitivity''
by Feldman and Cousins
\cite{Feldman-Cousins-98}
and we have argued that
this name is not appropriate because
this quantity does not represent
the capability of an experiment to reveal a signal.
We proposed to
call this quantity
\emph{``exclusion potential''}.

Considering the case of
a Poisson process with known background,
we have shown how the exclusion potential and the standard deviation
of the upper limit
in the absence of a signal
may help to choose the method
that is more appropriate for obtaining reliable upper limits
(Section~\ref{Exclusion}).
We have also defined the Pull of a null result,
that quantifies the reliability of an experimental upper limit.
In Section~\ref{Detection}
we have defined the upper and lower detection functions,
that give the most likely outcome of an experiment
if there is a signal.
In Section~\ref{Sensitivity} we proposed
an appropriate
definition of sensitivity of an experiment to a signal.
These definitions apply to both Frequentist and Bayesian
statistical theories and
can be easily generalized to any process
in which a quantity $\mu$
with known probability distribution is measured:
the upper (lower) detection function is the
average of the upper (lower) limit
and the $100\lambda\%$ sensitivity
is the lower value of $\mu$
for which there is a probability $\lambda$
to find a positive lower limit.

We considered explicitly
the case of a Poisson process with known background
in the framework of Frequentist and Bayesian
statistical theories,
but similar considerations and conclusions apply also to the case
of a Gaussian distribution with boundary.

\newpage

\begin{thebibliography}{10}

\bibitem{Feldman-Cousins-98}
G.J. Feldman and R.D. Cousins,
Phys. Rev. D \textbf{57}, 3873 (1998), physics/9711021.

\bibitem{Neyman-37}
Philos. Trans. R. Soc. London Sect. A \textbf{236}, 333 (1937),
reprinted in
\textit{A selection of Early Statistical Papers on J. Neyman},
University of California, Berkeley, 1967, p.~250.

\bibitem{Eadie-71}
W.T. Eadie, D. Drijard, F.E. James, M. Roos and B. Sadoulet,
\textit{Statistical Methods in Experimental Physics},
North Holland, Amsterdam, 1971.

\bibitem{Kendall-2A}
A. Stuart, J.K. Ord and S. Arnold,
\textit{Kendall's Advanced Theory of Statistics},
Vol.~2A,
\textit{Classical Inference \& the Linear Model},
Sixth Edition, Halsted Press, 1999.

\bibitem{Cousins-95}
R.D. Cousins,
Am. J. Phys. \textbf{63}, 398 (1995).

\bibitem{Giunti-bo-99}
C. Giunti,
Phys. Rev. D \textbf{59}, 053001 (1999), hep-ph/9808240.

\bibitem{Ciampolillo-98}
S. Ciampolillo,
Il Nuovo Cimento A \textbf{111}, 1415 (1998).

\bibitem{Roe-Woodroofe-99}
B.P. Roe and  M.B. Woodroofe, 
Phys. Rev. D \textbf{60}, 053009 (1999), physics/9812036.

\bibitem{Mandelkern-Schultz-99}
M. Mandelkern and J. Schultz,
J. Math. Phys. \textbf{41}, 5701 (2000), hep-ex/9910041.

\bibitem{Punzi-99}
G. Punzi, hep-ex/9912048.

\bibitem{CLW-2000}
Proc. of the CERN
Workshop on ``Confidence Limits'',
CERN, 17-18 Jan. 2000,
edited by F. James, L. Lyons and Y. Perrin,
CERN 2000-005.

\bibitem{Giunti-Laveder-power-00}
C. Giunti and M. Laveder,
hep-ex/0011069.

\bibitem{PDG-98}
C. Caso \textit{et al.},
Eur. Phys. J. C \textbf{3}, 1 (1998).

\bibitem{DAgostini-YR3-99}
G. D'Agostini,
CERN Yellow Report 99-03
(available at
http://{\-}www-{\-}zeus.{\-}roma1.{\-}infn.{\-}it/{\-}\~{}agostini/{\-}prob+{\-}stat.{\-}html).

\bibitem{Baudis-99}
L. Baudis \textit{et al.},
Phys. Rev. Lett. \textbf{83}, 41 (1999), hep-ex/9902014.

\bibitem{Simkovic-99}
F. Simkovic \textit{et al.},
Phys. Rev. \textbf{C60}, 055502 (1999), hep-ph/9905509.

\bibitem{KARMEN-Moriond99}
T.E. Jannakos (KARMEN Coll.), hep-ex/9908043.

\bibitem{Hernandez-Nava-Rebecchi-discovery-96}
J.J. Hernandez, S. Nava and P. Rebecchi,
Nucl. Instrum. Meth. \textbf{A372}, 293 (1996).

\end{thebibliography}
\end{document}